# Cryogenic Safety Aspect of the Low-β Magnet Systems at the Large Hadron Collider (LHC)

\* Darve C.

\* Fermi National Accelerator Laboratory, Accelerator Division, Batavia, IL, USA

The low-β magnet systems are located in the LHC insertion regions around the four interaction points. They are the key elements in the beams focusing/defocusing process and will allow proton collisions at a luminosity of up to $10^{34} cm^{-2} s^{-1}$. Large radiation dose deposited at the proximity of the beam collisions dictate stringent requirements for the design and operation of the systems. The hardware commissioning phase of the LHC was completed in the winter of 2010 and permitted to validate this system safe operation. This paper presents the analysis used to qualify and quantify the safe operation of the low-β magnet systems in the Large Hadron Collider (LHC) for the first years of operation.

## INTRODUCTION

The eight low-β magnet systems are located at the Interaction Regions (IR), 2, 5 and 8. They are mainly composed of four quadrupole magnets (Q1, Q2a, Q2b, Q3) collectively called the inner triplet, corrector magnets and the electrical feed-boxes (DFBX). A cold beam separation dipole magnet (D1) is added to this system at the low luminosity points IP2/IP8. Figure 1 shows the layout of the IP1/IP5 system, including quadrupoles, corrector magnets, copper collimator (TAS), Beam Position Monitor (BPM) and DFBX.

Using an underground cryogenic system could put at risk personnel and equipment. This hazardous environment requires stringent hazard analyzes, e.g. oxygen deficiency and impact on equipment in case of helium vessel rupture. The definition of the Maximum Credible Incident (MCI) is introduced for this purpose.

Due to the proximity of the interaction point to the low-β magnets, the environment of these magnets is highly radioactive [1-3], which ultimately could compromise possible modifications of the low-β magnet systems after the first years of operation.

In the sake of providing a coherent and methodological approach across HEP laboratories, a systematic safety analysis is recommended for future projects.

This document aims at recalling information in a format that serves the purpose of the LHC cryogenic safety aspect of the low-β magnet systems. For that matter, we recall the different CERN tools used to manage the equipment database and the guidance for the engineering risk assessment approach.

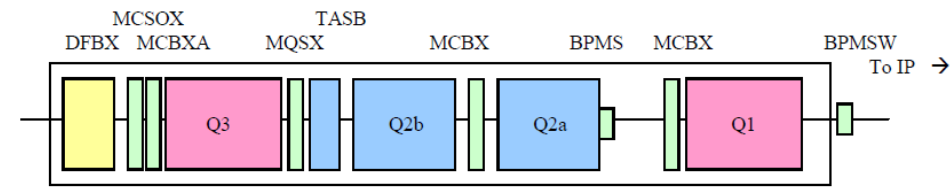

Figure 1 Layout of the IP1/IP5 low-β magnet system

# DESIGN OF THE LOW-β SYSTEM MAGNET

The cryogenic safety approach includes system specifications, system thermo-mechanical design, Process and Instrumentation Diagrams (P&ID), instrumentation choice, engineering drawings, spare and replacement part requirements, test acceptance criteria and interlock safety systems. These aspects were integrated within the LHC cryogenic framework and control process activities.

The typical specification is the functional specification written by the project engineer, checked by the project manager of the US-LHC collaboration, and approved by CERN management [4]. The typical DFBX functional analysis is available at [5]. Additional specifications were written, e.g. interface specifications [6], Technical Design Handbook (TDH) [7]. Various topics were identified within the low-β magnet system project, including the magnet powering and the definition of its cryogenic parameters.

## The low-β magnet system safety specification

The design constraints to build and operate the low-β magnet systems were:
1. Keep the system operation and maintenance safe for personnel and for equipment.
2. The equipment, instrumentation and design shall comply with the CERN requirements.
3. Use the Maximum Credible Incident (MCI) to assess the cryogenic risk.
4. Use materials resistant to the irradiation rate permitting an estimated machine lifetime, even in the hottest spots, exceeding 7 years of operation at the baseline luminosity of $10^{34} cm^{-2} s^{-1}$.
5. Apply the ALARA principle (As Low As Reasonably Achievable).

# CRYOGENIC RISK: IDENTIFICATION, ASSESSMENT AND MITIGATION

## Identification of risk

Each low-β magnet system has a capacity of about 150-180 kg of liquid helium at 1.9 K and they are located 80-120 m underground. This required an elaborated safety analysis. The Maximum Credible Incident (MCI) is used to assess the risk to personnel and to equipment. It is also used to size the helium relief system. Following the September 19, 2008 incident, the MCI was redefined [8]. This worst case scenario is now defined as an electrical arc through the inner triplet bus bars at nominal current, of 7 kA for Q1, Q3 and 10.6 kA for Q2a and Q2b. This electrical arc would generate the rupture of the helium containment vessel, which would be accompanied by a pressurization of the insulating vacuum. For this first scenario, no personnel is present in the tunnel. The estimated size of the opening caused by this rupture is 60 $cm^2$. Note that the stored magnetic energy in the inner triplet (6.5 MJ) is almost as large as the stored energy in an LHC arc dipole (7.5 MJ).

The second scenario investigated ensures the safety of personnel. Indeed, personnel is allowed in the tunnel if the inner triplet magnets are electrically isolated or if they operate at reduced current Hence, we consider 1) a leak from the helium space to the insulating vacuum, and 2) an electrical arc through the inner triplet bus bars. This scenario results in an estimated opening of 4 $cm^2$ on the helium containment vessel.

## Consequences of the MCI

The oxygen deficiency hazard analysis of the low-β magnet system is scaled down from the arc cryo-magnet analyses [2]. The rupture of interconnect is taken into account, leading to a substantially enhance of the relief capacity. Indeed, bus splices are located inside thin-walled (0.25 mm) bellows, which are vulnerable to puncture in case of an electrical arc.
It completes the design analysis [9-10]. The MCI for the low-β magnet system gives an estimate of the mass flow relieved in the LHC tunnel.

The estimated helium mass-flow relieved through the safety relief devices is on the order of 15 kg/s. The assessment of the risk due to the MCI can be listed as follows (Table 1).

Table 1 Assessment of the risk due to the MCI

| Case 1: Electrical arc through the inner triplet conductor at nominal current (no personnel is allowed in the tunnel). |
|---|
| • Opening to the vacuum/helium space = 60 cm$^2$ <br> • Maximum pressure in the insulating vacuum shall not exceed 1.17 bara <br> • Maximum flow venting at the safety relief device = 15 kg/s <br> • Helium discharge temperature though the safety relief valve = 20 K <br> • Number of recommended safety relief device = 3 DN200 + 3 DN65 |
| Case 2: Minor electrical arc through the inner triplet conductor at reduced current or leak from the helium space to the insulating vacuum (personnel is allowed in the tunnel). |
| • Opening to the vacuum/helium space = 4 cm$^2$ <br> • Maximum pressure in the insulating vacuum shall not exceed 1.03 bara <br> • Maximum flow venting at the safety relief device = 1 kg/s <br> • Helium discharge temperature though the safety relief valve = 80 K <br> • Number of recommended safety relief device = 1 DN200 (use a staged relief, with one dedicated relief devices opening at a lower pressure level than the others) |

Hardware implementation to mitigate a possible helium pipe rupture

Pressure safety relief devices insure that the eight liquid helium vessel and its vacuum vessel remain below 2.5 barg and 0.15 barg, respectively. Quench valves vent helium flow out of the LHe triplets' magnet.

The original low-β magnet system cryostat was equipped with three DN65 parallel plates. To address the MCI consequences, three additional DN200 safety relief devices were installed at the bottom of the cryostat, at the interface between the different quadrupole magnets of the inner triplet. They are designed to open in a staged manner. A ducting device (deflectors) is considered to route the helium gas away from the low-β working area and away from the escape door. This implementation will 1) allow personnel to work on the low-β system when the magnet energy is reduced and 2) will protect carbon steel equipment. Deflectors are expected to be installed during the long shut-down 2011.

Oxygen Deficiency Hazard (ODH) monitoring systems, signs, evacuation sirens and flashing light are installed in the tunnel to mitigate risk; they are redundant in critical areas such as the low-β magnet system.

## RADIOLOGICAL RISK: IDENTIFICATION, ASSESSEMENT AND MITIGATION

Identification of risk

The low-β magnet system high-radiation environment presents a continuous risk to personnel and to equipment. Ionization radiation is much higher in the magnet inner components closer to the beam. For comparison, the expected nominal LHC operation annual absorbed radiation dose at the arc magnet and at the high luminosity points, IP1/IP5, low-β regions is 1 and 1000 Gy on the vessel outside, respectively.

The radiation dose rate, at the nominal luminosity, estimated for the DFBX vacuum vessel at the high luminosity points, IP1/IP5, is of the order of 1 mGy/sec.

The power density, power dissipation, particle fluxes and spectra, accumulated dose and residual dose rate in the component of the low-β magnet system were studied [1, 11]. The low-β magnet system dimensions and materials were considered for the calculation. Table 2 summarizes the heat load $P$ and peak yearly dose $D$ in the IP1/IP5 DFBX elements. The life time of low-β magnet components were considered for the nominal luminosity of $10^{34}$ cm$^{-2}$s$^{-1}$ with regard to material properties.

Table 2 Heat load $P$ and peak yearly dose $D$ in the IP1/IP5 DFBX elements [2]

| Element | z-region (m) | P (W) | D (kGy/yr) |
|---|---|---|---|
| Pipe | | 0.841 | |
| Bore | | 1.994 | |
| Helium | 54.45-58.83 | 0.108 | 523.2 |
| Jack | | 0.936 | 310.6 |
| Ins+vessel | | 0.488 | |
| r=9 cm | | 1.014 | 74.18 |
| r=15 cm | 54.485-58.795 | 0.470 | 20.85 |
| r=30 cm | | 0.272 | 6.074 |

Radiological specification
The high-radiation environment requires analyzing the composition of the low-β magnet components w.r.t. the dose D distribution. Ionizing radiation can cause important damage to materials, by changing their mechanical properties, e.g. the radiation dose limit for bare Teflon at room temperature is 1 Mrad = 10 kGy. Hence, no Teflon is used at the proximity of the low-β magnet systems.

- Use only materials resistant to the irradiation rate above 1 mGy/sec, permitting an estimated machine lifetime, even in the hottest spots, exceeding 7 years of operation at the baseline luminosity of $10^{34}cm^{-2}s^{-1}$.
- Keep residual dose rates on the component outer surfaces of the cryostats below about 0.1 mSv/hr.
- The peak power density threshold, to the inner triplet, shall remains below the magnet quench limit of 1.6 mW/g. A safety factor of 3 is used in the calculation.
- Dynamic heat loads to the inner triplet shall not exceed 10 W/m.

Radiological risk mitigation
The life time of the low-β magnet components were considered for the nominal luminosity w.r.t. materials properties. The high dose estimate drives the mitigations as varied as:

- The inner-triplet final design included additional radiation shielding and copper collimator (TAS).
- Preference of PEEK versus Kel-F material used for the DFBX low temperature gas seal.
- The chosen instrumentation and equipment are radHard and halogen free [3].
- The process control makes use of different interlocks and alarm level for each operating mode [14].
- LHC tunnel accesses modes were defined, e.g. control and restricted modes.
- Procedures are being written based on lessons learned and to limit the exposition time.
- Specific hazard analysis is requested to intervene on the low-β magnet systems.

**TOOLS FOR RISK ASSESSMENT AND IMPLEMENTATION**

Software implementation to control operation upsets
Alarm, software interlocks and hardware interlocks are used in the low-β system. Bases on the LHC hardware commissioning lessons learned, threshold of interlocks were updated. New interlocks were added to comply with the updated LHC requirements. A simplified interlock list can be listed as follows:

- The so-called "Cryo-Start" and "Cryo-Maintain" are software interlocks ensuring that the magnets are energized in good conditions [14].
- Temperature switches ultimately protect the operation of the HTS leads by means of the power converter.
- Interlocks on insulating vacuum measurement provoke the lost of the cryogenic conditions to power the magnets.
- DFBX current leads cannot be energized if the voltage drop is more than 160 mV.
- A rise in the helium distribution pressure will provoke the closure of valves to isolate the DFBX.

Personnel training
In addition to the use of software and hardware to limit risks, personnel's training is of prime importance. New classes comply with the CERN safety policy and train the personnel to behave safely in a cryogenic and radiation environment. Awareness and preventive actions are mandatory to complete each technical task. Dedicated hazard analyses are enforced to work in the low-β magnet system area.

Tools to assess the consequences
The identification of the equipment permits to perform a systematic Failure Mode and Effect Analysis (FMEA) e.g a cryo-heater located inside the helium cold mass can fail on, provoking the helium to boil-off. This failure mode is safe because the helium vessel safety relief valves are properly sized. In addition, redundancy was considered for the cryo-heater in order to improve the system availability.

The "What-Ifs" analysis is used to assess the risk and to improve the system design and its performance. Devices/components were checked for unsafe conditions arising from the loss of electrical power, loss of instrument air causing component to fail close or open, or erroneous value measuring. The considered scenarios are shown in Table 3 (consequences and conclusions are not shown):

Table 3 Example of a simplified "What-Ifs" Analysis

| Quench on the low-β magnet system | Power Supply Power Outage |
|---|---|
| Cold compressor stops | Thermometry crate dies |
| Compressed air fails | Fieldbus: e.g. Profibus or WorldFip fails |
| Cryostat Insulating vacuum break | Industrial PC: e.g. FEC fails |
| QRL line rupture | PLC fails |
| Helium return line leaks/ruptures | Ethernet Network fails |
| He supply line leaks/ruptures | UNICOS/SCADA communication loss |
| Water cable leaks/ruptures | CIET communication loss |
| Current leads overloaded | DB, Logging communication loss |
| Beam Interlock System Fails | QPS and power supply fail |
| Large radiation dose achieved | Power Interlock Controller fails |

Tools for data management

The equipment properties are recorded using the CERN Management Equipment Folder (MTF) [12]. This manufacturing- and operation-oriented tool lists the relevant equipment using an assembly breakdown structure [3]. Individual equipment and instrumentation data like calibration and specifications are made available. None-conformity and radiation survey data are reported for the given equipment.

The second project-oriented CERN tool, which provides a common access point for all engineering documents, is the Engineering Data Management System (EDMS) [13]. This share point also houses the different elements of the risk assessment. The specifications and changes are accessible to collaboration members throughout the project. The EDMS tool has been used to achieve and control the different documents throughout the process.

**OPENING TO A NEW ENGINEERING PROCESS APPROACH**

Since engineering is human and risk needs to be assessed for every project, a new engineering manual was issued at Fermilab [15]. This risk-based graded approach provides safe, cost-effective and reliable designs. Figure 2 shows the Fermilab Engineering Process sequences. The implementation of this engineering process shall be sufficiently flexible to loop within the given sequences. The implementation of this process will be adjusted to the Fermilab future projects.

This process is accompanied with a new Fermilab lab-wide tool to archive and control engineering documents, so-called, TC/ Electronic Document Management System (EDMS).

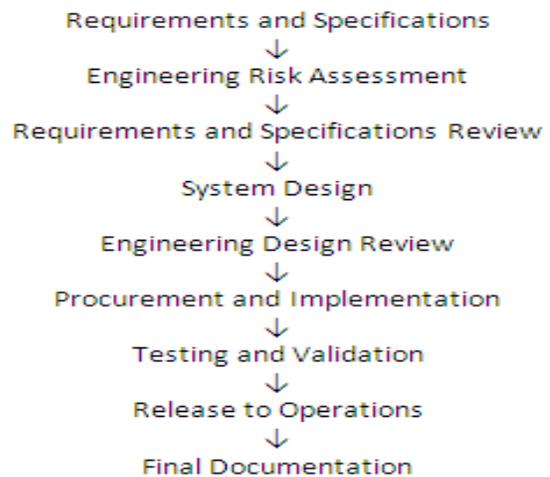

Figure 2 Example of Engineering Process [15]

# CONCLUSION

A simplified risk assessment has been conducted for the low-β magnet systems design and installation. We reviewed the different steps used for the cryogenic safety aspect of the low-β magnet system, with emphasize on several key issues. The FMEA and "What-Ifs" analysis technique examines the consequences of system failures and upsets. It also permits to prevent the procedural errors. Following the LHC hardware commissioning, the instrument list and hazard analysis were updated.
   The system reliability, availability and maintainability have been considered.
   A sound engineering process was used and lessons learned can be applied to future projects.

# ACKNOWLEDGEMENT

The author would like to thank the TE/CRG personnel, the integration group (ILC), the safety group (TGS) and the hardware commissioning team for their technical support. Thanks to Nikolai Mokhov, Laurent Tavian, Tom Nicol and Jim Strait for sharing their expertise. Contributions from Herve Prin have permitted to install safety relief devices.